\documentstyle[epsf]{article}

\begin{document}
\baselineskip8mm
\title{The generality of inflation in some closed FRW models
with a scalar field}
\author{S. A. Pavluchenko$^{1}$, \ N. Yu. Savchenko$^{2}$ \
   and \ A. V. Toporensky$^{1}$}
\date{}
\maketitle
\hspace{-6mm}
$^{1}${\em Sternberg
Astronomical Institute, Moscow State University, Moscow 119899, Russia}\\
$^{2}${\em Quantum Statistics and Field Theory Department, Physical Faculty,
 Moscow State University,
Moscow 119899, Russia}

\begin{abstract}
The generality of inflation in closed FRW Universe is studied for the
models with a scalar field on a brane and with a complex scalar field.
The results obtained are compared with the previously known results for
the model with a scalar field and a perfect fluid. The influence of the
measure chosen in the initial condition space on the ratio of inflationary
solution is described.
\end{abstract}

\section{Introduction}
In the recent two decades the dynamics of an isotropic Universe filled with
a massive scalar field attracted a great attention. There are
several possible dynamical regimes in this system \cite{star}.
From the physical point of view the inflationary regime
is the most important one \cite{Linde},
therefore, one
of the problems of mathematical cosmology is to describe the set of initial
data leading to inflation. This problem for a massive scalar field
was studied in \cite{four,B-Kh}. The main result of these studies is that
the set of initial conditions required for inflation strongly depends on
the sign of the spatial curvature.
If it is negative or zero, the scale factor of the
Universe $a$ cannot pass through
extremum points. In this case all the
trajectories starting from
a sufficiently large initial value of the
scalar field $\varphi_0$, reach a slow-roll regime and
experience
inflation. If we start from the Planck energy, a measure of
non-inflating trajectories for a scalar field with
the mass $m$ is about $m/m_{P}$.
From observational
reasons, this ratio is about $10^{-5}$
so almost all trajectories lead to the inflationary regime. However,
positive spatial curvature allows a trajectory to have a point of maximal
expansion which  results in increasing the measure of non-inflating
trajectories. In a famous angular parametrization described below
this measure is about $\sim 0.3$ \cite{B-Kh}.

Hence, the sign of the spatial curvature, which is rapidly "forgotten" after
the inflationary regime established, is very important during a possible
pre-inflationary period. Another important characteristic of the
pre-in\-fla\-ti\-on\-ary era which
is also `forgotten" during inflation is the possible presence of a
hydrodynamical matter in addition to the scalar field. Decreasing even more
rapidly than the curvature with increasing scale factor $a$, the energy
density of hydrodynamical matter could not affect the slow-rolling
conditions and so has a tiny effect on the dynamics if the spatial
curvature
is non-positive.  However, the conditions for extrema of $a$ can change
significantly in a closed model, so this model requires special analysis.
It was done in \cite{topor} where the maximum measure of non-inflationary
trajectories in the angular parametrization is found to be $ \sim 60\%$
with the significant dependence of this measure on the density of
hydrodynamical matter.

In recent years some interesting modification of standard scenario
become the matter of investigations. After a famous paper of Randall
and Sundrum \cite{R-S} a
 lot of effort was done on the idea that
our Universe is a boundary (brane) of a manifold with a larger dimensions (bulk).
Inflation caused by a massive scalar field on the brane in the case
of flat FRW brane metric was considered in \cite{W-M}. One of the
principal features of the brane scenario is the presence of the
so called "dark radiation" term in the equations of motion.
In the case of FRW brane this term
is inversely proportional to the fourth power of the scale factor
(as an ordinary radiation matter),
so, it rapidly decreases during inflation. On the other hand, we
can expect that this term can alter the conditions for inflation
on closed FRW brane in a way similar to studied in \cite{topor}.
This problem is the topic of Sec.2 of the present paper.

Apart from branes, some generalizations of simple scalar field scenario
appeared also in the framework of "standard" cosmology. In this paper we
will consider one of them, namely a massive complex scalar field.
It is naturally appears in supersymmetry theories.
Some problems of quantum cosmology with a complex scalar field were
studied in \cite{Khalat,we}. Recent  proposals to use this type of scalar
field to explain dark matter and quintessence were put forward
in \cite{Tw,Kamion}. We will apply our analysis of generality
of inflation to the complex scalar field in Sec.3.

The results of studies of these three models - a scalar field with
a hydrodynamical matter, a scalar field on the brane and a complex
scalar field are compared in Sec.4.

\section {A massive scalar field on the brane}

In this section we consider a "three--brane universe" in a five dimensional (bulk)
space-time.
We will consider only purely nonstandard
brane dynamics when the matter density on the brane dominates the brane
tension.
Then the equations describing the cosmological
evolution of a closed FRW brane are ~\cite{Bin,Maeda}
\begin{equation}
\begin{array}{c}
\label{from}
\displaystyle \frac{\dot a^2}{a^2}=\frac{\kappa^4}{36}\rho^4+\frac{C}{a^4}-
     \frac{1}{a^2},  \lower 6mm \hbox{}  \\
\displaystyle \frac{\dot a^2}{a^2}+\frac{\ddot a}{a}=
       -\frac{\kappa^4}{36}\rho(\rho+3p)-\frac{1}{a^2}.
\end{array}
\end{equation}

Here $\kappa^2=8 \pi/m_{5P}^3$, $m_{5P}$ is the 5-dimensional fundamental Planck mass which can
differ significantly from 4-dimensional Planck mass observing on
the brane. For a review of the current status of the brane model
see, for example \cite{Maartens}.

The equations of motion for a brane universe with a massive scalar
field have the form


\begin{equation}
\ddot\varphi+3\frac{\dot a}a\dot\varphi+m^2\varphi=0,
\end{equation}

\begin{equation}
\frac{\ddot a}a+2\frac{\dot a^2}{a^2}+\frac2{a^2}= \frac{4 \pi^2}{3 m_{5P}^6}\left(m^4\varphi^4-\dot\varphi^4\right)+\frac{C}{a^4}
\end{equation}

with a constraint
\begin{equation}
\label{co}
\displaystyle \dot a^2+1=\frac{16 \pi^2}{9 m_{5P}^6}a^2\left(\frac12 \dot\varphi^2+
          \frac12 m^2\varphi^2\right)^2+\frac C{a^2}.
\end{equation}

The equations of motion in brane scenario in comparison with those
in standard cosmology characterizes by
quadratic dependence on energy-momentum tensor
and nonlocal effects from the free gravitational field in the bulk,
manifesting itself in the presence of a "dark radiation" term (the last
term in righthand side of Eqs. (3) and (4)).

The inflationary dynamics on the brane for the flat brane metric was
considered in Ref. \cite{W-M}. The situation appears to be not so
different from  the standard case, though the slow-roll regime is
changed.  The slow-roll condition
being $\varphi > m_{5P}$ in standard cosmology takes the form
$\varphi > m_{5P}/\sqrt{m/m_{5P}}$ in the brane model
and, hence, the measure of non-inflationary initial
conditions is $\sqrt{m/m_{5P}}$ instead of $m/m_P$ in standard scenario.
Our goal here is to study the positive curvature case.

First we need to specify a set
of initial conditions. It is common to use the hypersurface with  the
energy density equal to the Planck one (called the Planck boundary) as
the initial-condition space.
As was pointed out in Ref.\,\cite{BKH}
this choice is based on the physical considerations of inapplicability of
classical gravity beyond the Planck boundary and of the absence of
any information from this region. We remind the reader that in the brane scenario
the fundamental Planck scale limiting the applications of classical gravity
is different from the effective Planck scale on the brane.

To compare our results with those obtained in the standard scenario,
we rewrite the constraint equation so that the coefficient in the
lefthand side  be the same:
\begin{equation}
\frac{3 m_{5P}^2}{8 \pi} \left (\frac{\dot a^2}{a^2}+\frac{1}{a^2} \right )=
\frac{2 \pi}{3 m_{5P}^4} \left (\frac{\dot \varphi^2}{2}+\frac{m^2 \varphi^2}
{2} \right )^2 + \frac{3 m_{5P}^2}{8 \pi} \frac{C}{a^4}.
\end{equation}

Now, we start from the Planck boundary:
\begin{equation}
\label{Pl}
\frac{2 \pi}{3 m_{5P}^4} \left (\frac{\dot \varphi^2}{2}+\frac{m^2 \varphi^2}
{2} \right )^2 + \frac{3 m_{5P}^2}{8 \pi} \frac{C}{a^4} = m_{5P}^4.
\end{equation}

 To describe the initial data on the
Planck boundary
 it is useful to introduce
two dimensionless variables, as in \cite{B-Kh}.
One of them is the angular parameter $\phi$ defined as

\begin{equation}
\begin{array}{c}
\label{qq}
\displaystyle  \dot \varphi^2 = \left( m_{5P}^2 - \frac{3C}{8 \pi a^2} \right)^{\frac{1}{2}} \sqrt{\frac{6}{\pi}} m_{5P}^3 \cos^2 \phi \lower 6mm \hbox{}, \\
\displaystyle  m^2 {\varphi}^2 = \left( m_{5P}^2 - \frac{3C}{8 \pi a^2} \right)^{\frac{1}{2}} \sqrt{\frac{6}{\pi}} m_{5P}^3 \sin^2 \phi
\end{array}
\end{equation}
and the second one is
$z=\dot a/(a m)$. A couple $(z,\phi)$ determine the initial point on the
Planck boundary completely.
The variable $z$ can vary in a compact region
from $0$ to $z_{\max}=\sqrt{8 \pi/3}(m_{5P}/m)$;
the corresponding initial values of the scale factor $a$ varies from
$a_{min}=\sqrt{3/(8 \pi m_{5P}^2)}$ to $+\infty$.

\begin{figure}
\epsfxsize=15cm
\centerline{{\epsfbox{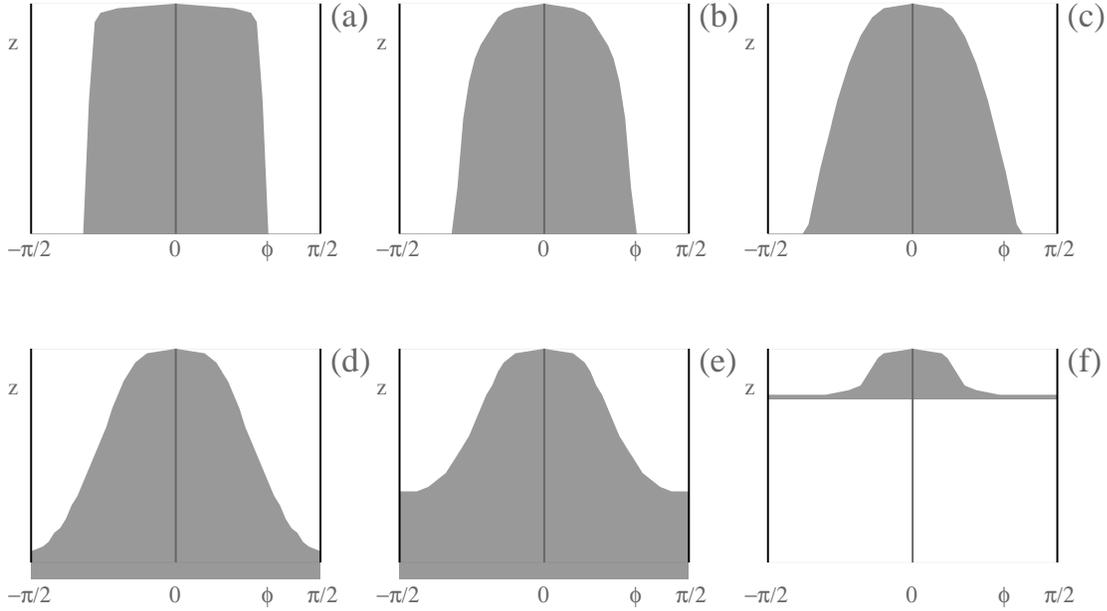}}}
\caption
{The area of initial data on the Planck boundary which do not lead
to inflation (gray). Fig.\,1(b) corresponds to the situation without
a "dark radiation"
term, other plots correspond to densities of the "dark radiation"
increasing  from Fig.\,1(a) to Fig.\,1(f) (see details in the text). The
value of $\phi$ varies is the range $ - \pi / 2 \le \phi \le \pi /2$ and the value of
$z$ in the range $0 \le z \le z_{max}$.} Initial data below the horizontal line
in Fig.1(f) are physically inadmissible.
\end{figure}

The numerical calculation results are plotted in Fig.1. The region where
inflation is possible is the gray one. The resulting measure of initial
data leading to inflation as a function of $C$ is plotted in Fig.2.
 Though this measure does not
change drastically, the configuration of corresponding regions
in initial condition space shown in Fig.1
undergoes substantial transformation. In particular, for large
enough $C$ inflation becomes impossible if the value of
initial Hubble parameter $H=\dot a/a$ is less than some critical one (or,
equivalently, initial spatial curvature is greater than some critical value).
We can easily
find analytically the minimal value of $C$ when this situation takes
place, i.e. when there are no inflation with initial $z=0$. To do this
we remark from Eq.(4) that if we start with some initial energy $E$
with $H=0$ (i.e. in the extremum of $a$),
then  possible initial $\varphi$ lie in the interval
\begin{equation}
\varphi^2<\frac{3 m_{5P}^3}{2 \pi m^2 a_{in}} \left (1-
\frac{C}{a_{in}^2} \right)^{1/2},
\end{equation}
where initial scale factor $a_{in}=\sqrt{3 m_{5P}^2/8 \pi E}$.

From the other side, the direction of trajectory in the extremum of $a$
depends on the sign of the second derivative $\ddot a$.
 Substituting $\dot a=0$ into Eq. (3) and expressing
$\dot \varphi$ from the constraint equation (4), we obtain that $\ddot a$
at the point $\dot a=0$ is
\begin{equation}
\ddot a=\frac{4 C}{a^3} - \frac{5}{a} +m^2 \varphi^2 \frac{4 \pi}
{m_{5P}^3} \left(1-\frac{C}{a^2} \right)^{1/2}.
\end{equation}

We can see that for initial $z=0$,
$\ddot a <0$ (Universe collapses  directly from the initial point)
when
\begin{equation}
\varphi^2< \frac{ m_{5P}^3}{4 \pi m^2 a_{in}}
\frac{5-4C/a_{in}^2}{(1-C/a_{in}^2)^{1/2}}.
\end{equation}

The intervals (8) and (10) coincide at the critical value of initial scale factor
$a_{in}=a_{cr}=\sqrt{2 C}$.
It means that if we start from the Planck boundary,
then  for $C=C_1=3/(16 \pi m_{5P}^2)$ (in this case
$a_{cr}=a_{min}$) all trajectories with $z=0$ go to singularity
directly from
the initial point and can not reach inflation.
For $C>C_1$ it is impossible to reach inflation from some high-curvature
initial data independently on the initial $\phi$.

The curve (8) does not exist for $a_{in}<\sqrt{C}$.
 If initial $a$ corresponding to
a given initial $z$ is less than this value then
it is impossible to find any $\phi$,
such that the total energy of the system ("dark radiation" plus the
scalar field) does not exceed the Planck energy.
The Planck boundary is now only part of the initial rectangular area
in the coordinates $z$ and $\phi$ (see Fig.4). This situation occurs
if $C>C_2=3/(8 \pi m_{5P}^2)$.
 The ratio of non-inflation trajectories reaches its maximum
at $C=C_2$ (the maximum value of this ratio is $65 \%$)
and steadily decreases for larger $C$.

\begin{figure}
\epsfxsize=10cm
\centerline{{\epsfbox{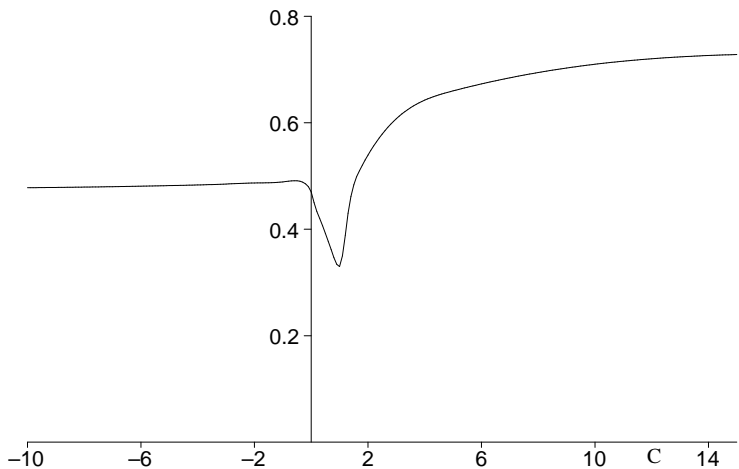}}}
\caption
{The ratio of inflationary solutions in the angular parametrization
as a function of the "dark radiation" density. The "dark radiation"
parameter $C$ is measured in the units of $3/8 \pi m_{5P}$.}
\end{figure}

We also checked numerically, that another choices of initial energy
$E$ lead to qualitatively the same picture with appropriate rescaling
of the parameter $C$.

A negative $C$ (Fig.1(a)) does not change the situation significantly in comparison with the case $C=0$ 
(Fig.1(b)). The results for positive $C$ are similar to those received in~\cite{topor} where the
degree of generality of inflation in ordinary FRW Universe
with massive scalar field and hydrodynamical matter
is analyzed.
It is natural, since the $C$-containing term behaves exactly as a
hydrodynamical matter with the equation of state $p=\epsilon/3$.

\section{A complex scalar field}

\subsection{Equations of motion}
In this section we return to standard FRW cosmology but consider a more
complicated version of the scalar field, namely, the massive complex scalar field.
The action for this model is
\begin{equation}
S=\int d^4 x \sqrt{-g}\left( \frac{m_P^2}{16\pi} R +\frac{1}{2}
{g^{\mu\nu}}{{\phi}_{\mu}^{*}}{\phi}_{\nu} -\frac{1}{2}
{m^2}\phi {\phi}^{*} \right) 
\end{equation}

and now the Planck mass $m_P$ has its common sense.

We will use a suitable representation of the complex scalar field \cite{Khalat}
$$
\phi=x \exp(i\theta),
$$
where $ x $ is absolute value of the complex scalar field and $ \theta $ is its
phase.

Using this representation we can see that the
phase variable $ \theta $ is cyclical and its conjugate momentum $ {p_{\theta}}
 $ is
conserved. We call it as a "charge" of the Universe $ Q $:
\begin{equation}
p_{\theta}=Q={a^3}{x^2}{\dot \theta}.
\end{equation}

Expressing the phase $\theta$ through the charge $Q$
we can get the following equations of motion \cite{Khalat}:
\begin{equation}
\frac{m_P^2}{16\pi}\left(\ddot a+\frac{{\dot a}^2}{2a}+\frac{1}{2a} \right) +\frac{a{\dot x}^2}{8}
-\frac{m^2 x^2 a}{8}+\frac{Q^2}{4a^5 x^2} =0,
\end{equation}

\begin{equation}
\ddot x +\frac{3 \dot x \dot a}{a}+m^2 x - \frac{2 Q^2}{a^6 x^3} =0.
\end{equation}

Besides we can write down the first integral:

\begin{equation}
\frac{3 m_P^2}{8\pi}\left(H^2+\frac{1}{a^2}\right)=\frac{m^2 x^2}{2}+\frac{{\dot x}^2}{2}+
\frac{Q^2}{a^6 x^2}.
\end{equation}




For future use we also find a curve, separating possible 
maxima and minima of the scale factor. Following the same procedure, as
in the previous section we can express $\dot x$ from the constraint (15)
and substitute it in Eq. (13). This equation gives now for $\dot a=0$

\begin{equation}
\ddot a = \frac{4 \pi m^2 x^2 a}{m_P^2} - \frac{2}{a}.
\end{equation}

So, the separating curve is
\begin{equation}
x^2 \le \frac{1}{2 \pi} \frac{m_P^2}{m^2 a^2}
\end{equation}
and does not depend on the charge $Q$ \cite{we}.

\subsection{Inflation in models with a complex scalar field}

As in the previous section, we start from the Planck boundary
\begin{equation}
\frac{3{m_P^2}}{8\pi}\left( H^2 +\frac{1}{a^2} \right)  = \frac{m^2 x^2}{2}+ \frac{{\dot x}^2}{2} +
\frac{Q^2}{a^6 x^2}  ={m_P^4}.
\end{equation}

Then it is necessary to introduce a measure on the initial condition space.
We remind the reader that in the case of a real scalar field the usual way to parameterize
the Planck boundary is to introduce two dimensionless variables,
$z= H/m$ and $\phi$ via the angular parametrization, which in standard
cosmology has the form

$$
\frac{m^2 {\varphi}^2}{2} + \frac{ {\dot \varphi}^2 }{2} = m_P^4; \quad \frac{m^2 {\varphi}^2}{2} = m_P^4 \sin^2 \phi;
\quad \frac{ {\dot \varphi}^2 }{2} = m_P^4 \cos^2 \phi.
$$

However, in our case we have additional term in the left part of (18),
so we can not
use this way. Instead, the second dimensionless coordinate we chose as
$X=x/m_{P}$.

 Two variables, $z$ and $X$,
completely determine initial conditions on the Planck boundary. We can
see from Eq.(15) that possible initial $X$ for a given $z$ lie in the
interval $[X_{min},X_{max}]$, where

$$
X_{min} = \frac{m_P^2}{m^2} \left(1 - \sqrt{1-\frac{2 m^2 Q^2}{m_P^8}\left(\frac
{8 \pi m_{P}^2}{3}-z^2 m^2\right)^3} \right)^{1/2},
$$

$$
X_{max} = \frac{m_P^2}{m^2} \left(1 + \sqrt{1-\frac{2 m^2 Q^2}{m_P^8}\left(\frac
{8 \pi m_{P}^2}{3}-z^2 m^2\right)^3} \right)^{1/2}.
$$
The variable $z$ as in the previous section varies from $0$ to
$\sqrt{8 \pi/3}(m_P/m)$.

Numerically found initial conditions leading to inflation for a real scalar
field ($Q=0$) in the coordinates $z$ and $x/m_{P}$ are shown in Fig. 3 (a).
The measure of non-inflating trajectories is equal to $0.52$.
The difference between this value and $0.3$ found in
\cite{B-Kh} arises completely from differences in chosen measures.
  With growing $Q$ the picture experiences changes
similar to the case studied in the previous section.

\begin{figure}
\epsfxsize=15cm
\centerline{{\epsfbox{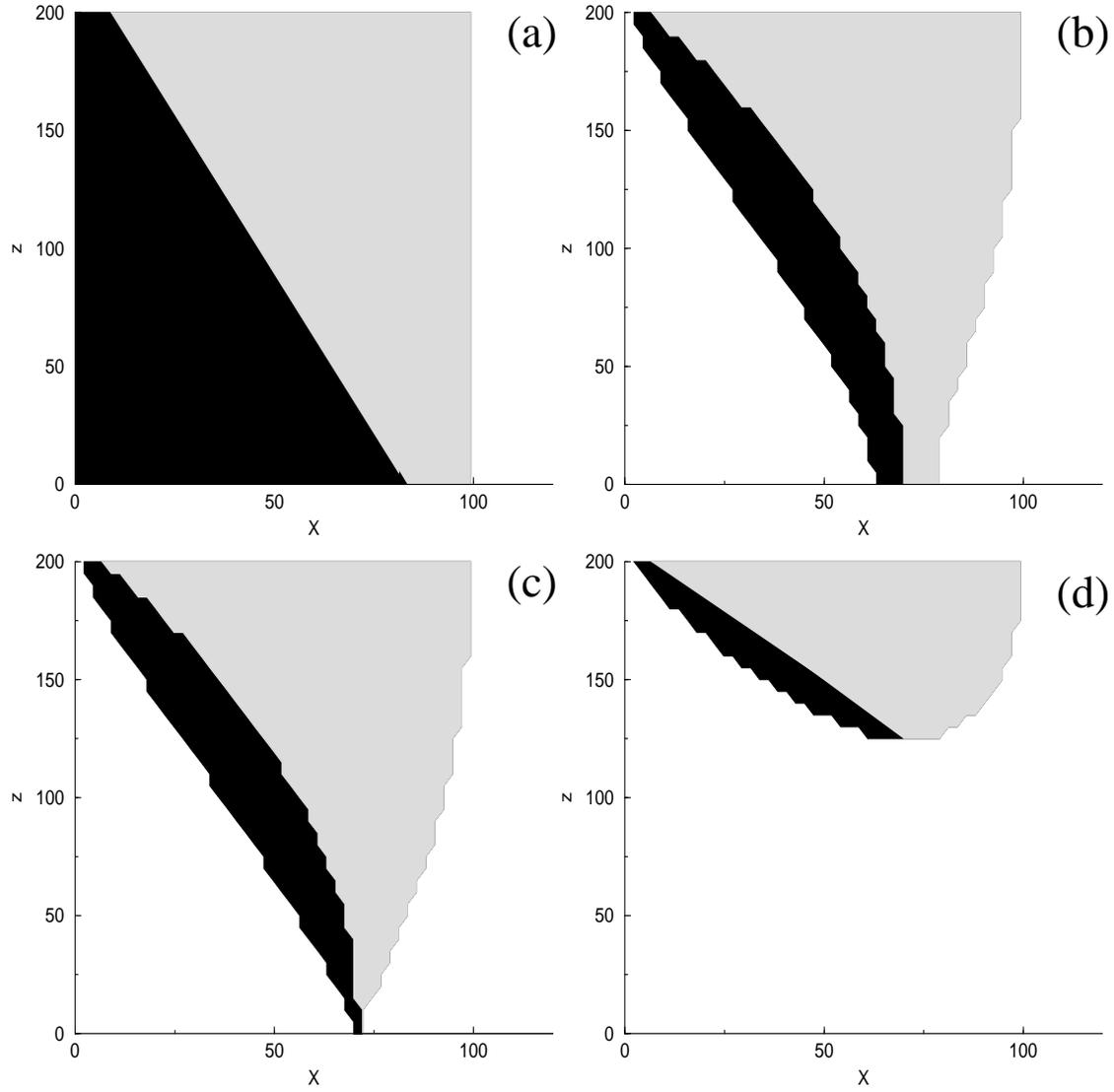}}}
\caption
{The area of initial data on the Planck boundary which does not lead
to inflation (black) and leads to inflation (gray).
 Fig.\,3(a) corresponds to the situation without
 charge,
 other plots correspond to charges
increasing  from Fig.\,3(b) to Fig.\,3(d).}
\end{figure}

If we start with zero $z$ and some initial $a$,
the maximum and minimum values of the scalar field $x$
lie on the curve
\begin{equation}
\frac{3{m_P^2}}{8\pi a^2} = \frac{m^2 x^2}{2}+
\frac{Q^2}{a^6 x^2}.
\end{equation}

The separating curve is given by Eq. (17)
and crosses the curve (19) at

\begin{equation}
a^2=\frac{8 \pi^2 Q^2 m^2}{m_{P}^4}.
\end{equation}

Remembering that the initial value of the scale factor,
corresponding to $z=0$ on the Planck boundary is
$a_{min}=\sqrt{3/8 \pi m_{P}^2}$,
we find that for $Q=Q_1=(1/8 \pi) \sqrt{3/2{\pi}} (m_P/m)$
all trajectories with $z=0$
fall into a singularity. For $Q>Q_1$ there is a high-curvature region
with no way to reach inflation (Fig.3(c)).

As in the previous section, there exists some value of charge
for which start from $z=0$ becomes impossible.  It happens if
$Q$ exceeds $Q_2=(3/32 \pi) \sqrt{{3}/{ \pi}} ({m_{P}}/{m})$ (Fig.3(d)).

It is easy to see, that these values are rather close to each other:

\begin{equation}
\frac{Q_2^2}{Q_1^2} = \frac{9}{8}.
\end{equation}

\begin{figure}
\epsfxsize=10cm
\centerline{{\epsfbox{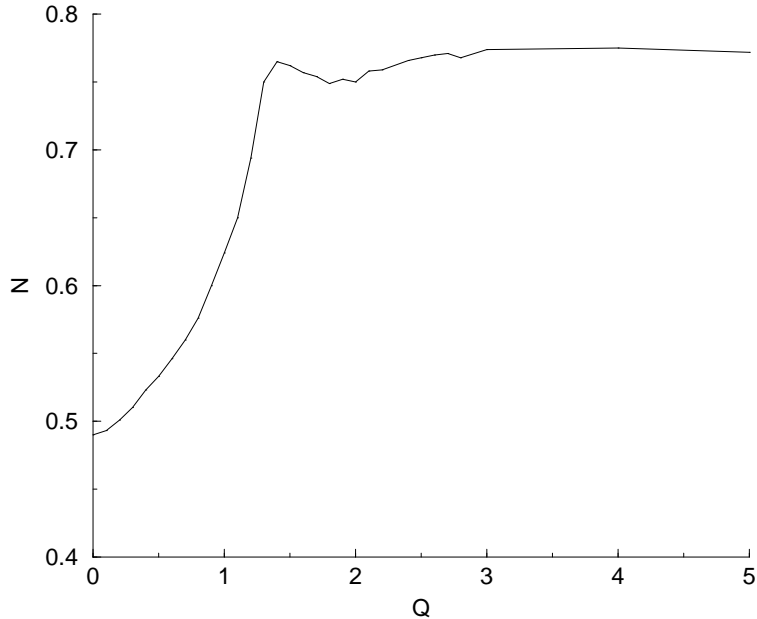}}}
\caption
{The ratio of inflationary solutions N
as a function of charge Q. Scalar field mass \qquad \qquad
$m = 1.4 \times 10^{-2} m_P$.}
\end{figure}

As $Q_1$ and $Q_2$ depend on $m$, this dependence appears also in the curve
$N(Q)$, where $N$ is the measure of inflationary solutions. An example
for $m=1.4 \times 10^{-2} m_P$ is plotted in Fig.4. For another
$\tilde m$ this curve is similar with rescaling $\tilde Q \to Q (m/ \tilde m)$.

\section{ Discussion}
In the present paper we have considered two particular examples of a scalar
field dynamics on the closed FRW background. Together with the model of
a massive scalar field with a hydrodynamical matter studied in \cite{topor}
these cases, arising from rather different physical models, show several
common features.

 First of all, the measure of initial conditions
leading to inflation remains large enough, not falling below $35\%$.
If we consider a detailed structure of initial conditions
for recollapsing universes, we find that depending on the parameter
of the model (which can be a charge, a "dark radiation" density or
a density of a hydrodynamical matter) there are two situation possible.
For small values of the parameter inflation can start from an arbitrary
large initial curvatures, or, equivalently, arbitrary small initial scale
factors (of course, within the Planck limit). For the parameter bigger
than the critical one all trajectories from the high-curvature initial
data go back to a singularity instead of inflation. These critical
values are $C_1={3}/{16 \pi m_{5P}^2}$ for a "dark radiation" in the
brane model and
$Q_1=({1}/{8 \pi}) \sqrt{{3}/{2 \pi}} ({m_{P}}/{m})$ for
a charge. In the case of the massive scalar field with a perfect fluid
which has the equation of state $p=\gamma \epsilon$
we can introduce a constant parameter $D$ which is related to energy
density $\epsilon$ as $D=\epsilon a^{3(\gamma+1)}$.
In this case $D_1=(\frac{3}{8 \pi})^{1+p/2} m_{P}^{2-p} 2^{-p/2}$,
where $p=1+3 \gamma$ \cite{topor}.

There is also the second critical value of the parameter which is
equal to $C_2={3}/{8 \pi m_{5P}^2}$ for the "dark radiation",
$Q_2=({3}/{32 \pi}) \sqrt{{3}/{ \pi}} ({m_{P}}/{m})$
 for the charge and
$D_2=(\frac{3}{8 \pi})^{1+p/2} m_{P}^{2-p}$
for the perfect fluid density (the latter value is taken
from \cite{topor}). For bigger value of the parameter the Planck
boundary covers only a part of the initial rectangular. The
coordinate $z$ varies in the interval $z_{min} < z < z_{max}$
with some nonzero $z_{min}$, corresponding to
some minimal possible initial scale factor $a$. Smaller initial $z$ would
correspond to
smaller initial $a$ and the energy density in "dark radiation", charge
or matter terms would exceed the Planck energy which makes these
initial conditions physically inadmissible. Numerical data show also that
if the parameter
exceeds the second critical value the described above range of initial
$z$ not leading to inflation independently on the second coordinate
continues to exist (see Fig.1(f)), though being narrower with increasing
value of the parameter.

Looking on the overall measure of the initial condition allowing inflation
we can see also a remarkable difference between the brane scenario (Fig.2)
and the complex scalar field (Fig.4). In the former case this measure
decreases significantly with increasing $C$ up to the second critical value
$C_2$,
though in the latter case the measure is almost monotonically increasing
function of $Q$. In the model of scalar field with a perfect fluid,
studied earlier, the situation is similar to the brane case. However,
this difference mostly appears because of the different choice of the
measure. We remind the reader that in the brane scenario and the perfect fluid case we
use the angular parametrization of the initial condition space, and
in the complex scalar field case the value $x/m_P$ is chosen for
the second coordinate (the first coordinate, $z$, is the same in all three
cases). It is possible to use the latter measure for another cases
under consideration. The results are presented in Fig.5. For the model
with a perfect fluid we can see a very small decrease (about $8\%$) of the measure 
in comparison with about $2$ times drop of the measure in the angular parametrization.
In the brane scenario
we can see a rather small decreasing of the measure with the minimum
at $C \sim 0.7 C_2$. In this point the measure is $ \sim 88 \%$ of the
value at $C=0$ in comparison with about $1.5$  times drop in
the angular parametrization.

\begin{figure}
\epsfysize=7cm
\epsfxsize=15cm
\centerline{{\epsfbox{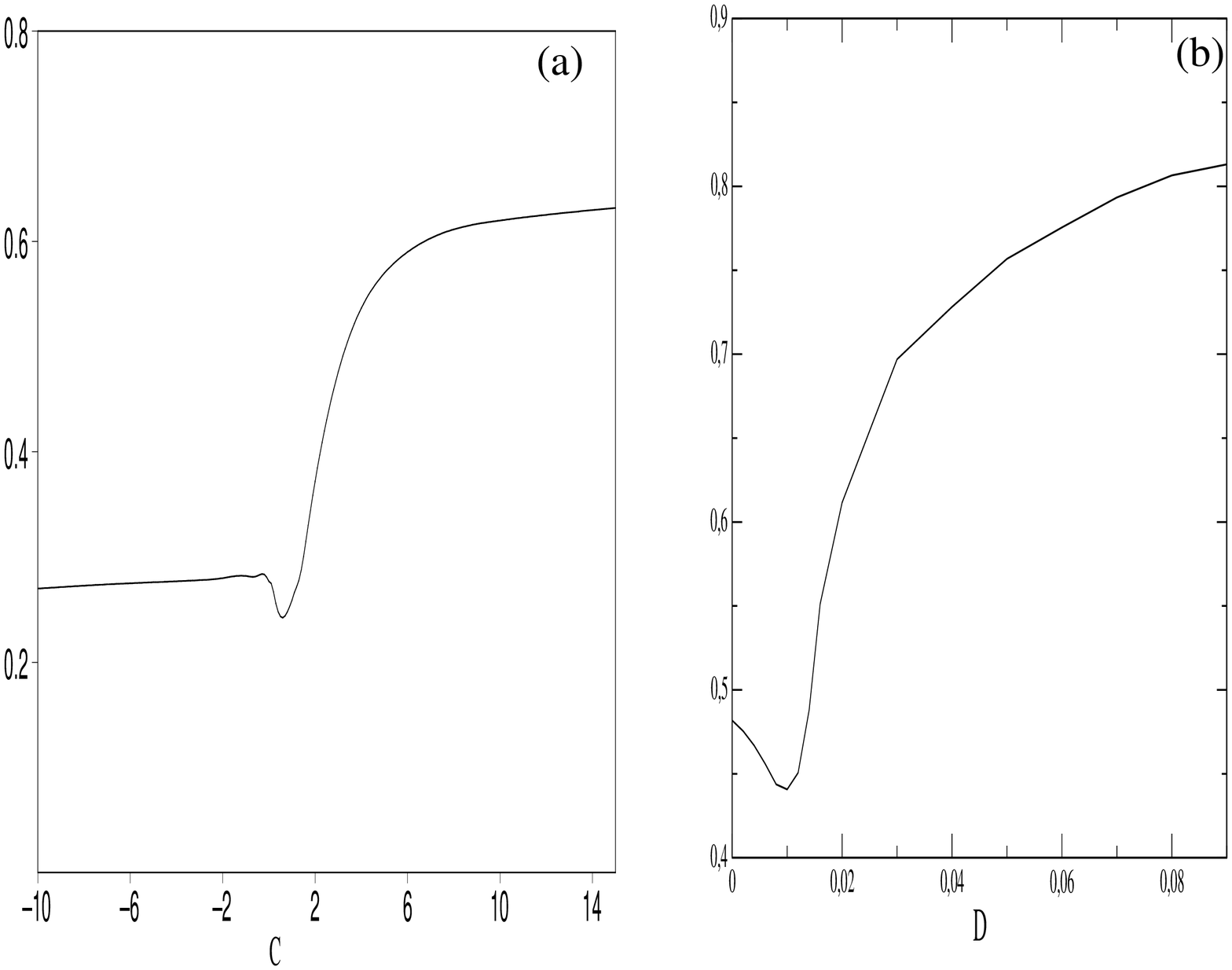}}}
\caption
{The ratio of inflationary solutions in the scalar field parametrization
as a function of the "dark radiation" density in the brane scenario (a)
and of the radiation matter density in the model with scalar field and
a perfect fluid (b). $C$ is measured in the units of $3/8 \pi m_{5P}$,
$D$ is dimensionless.}
\end{figure}

This difference can be easily understood by comparing of these two
parametrizations of the Planck boundary. The important feature
of the angular parametrizations  is that independently on the
possible initial $z$ on the Planck boundary the second
coordinate $\phi$ can vary in the same range $[-\pi /2; \pi /2]$. On the other
hand, the possible range for $\phi/m_P$ (or $x/m_P$ for the complex
scalar field) steadily decreases with increase of the initial curvature
(see Eq.(8)). So, the high-curvature initial data, less favorable
for inflation give bigger contribution for the measure based on
the angular parametrizations than for the measure based on $\phi/m_P$
as the second coordinate.

For parameter bigger than the second critical one the map from one
measure to another becomes even discontinuous. The line $z=z_{min},
\varphi \in [-\pi /2; \pi /2]$ in Fig.1(f) (consisting only on non-inflationary
trajectories) maps into a single point (compare Fig.1(f) and Fig.3(d)).
So, a definitely nonmonotonic behavior of the measure of inflationary
trajectories as a function of the parameter of the model in the case
of the angular parametrizations of the Planck boundary can be
explained as a measure effect.

\section*{Acknowledgments}
This  work  was  supported  via 'Universities of  Russia,  Fundamental
Investigation' grant No 990777 and partially  supported  by  Russian
Foundation   for   Basic  Research  via  grants  Nos  99-02-16224   and
00-15-96699.

\end{document}